# BEHAVIORAL ASPECTS OF SOCIAL NETWORK ANALYSIS

Sung Joo Park[1*], Jong Woo Kim[2], Hong Joo Lee[3], Hyun Jung Park[4],
Peter Gloor[5]

[1]KAIST Business School, [2]School of Business, Hanyang University, [3]Department of Business Administration, The Catholic University of Korea, [4]Department of Business Administration, Ewha Woman's University, Korea
[4]MIT Center for Collective Intelligence, U.S.A.
*Corresponding Author: sjpark@business.kaist.ac.kr

**ABSTRACT**

Contrary to the structural aspect of conventional social network analysis, a new method in behavioral analysis is proposed. We define behavioral measures including self-loops and multiple links and illustrate the behavioral analysis with the networks of Wikipedia editing. Behavioral social network analysis provides an explanation of human behavior that may be further extended to the explanation of culture through social phenomena.

**INTRODUCTION**

Although social network analysis methods have been progressing remarkably in the last decade, there are methodological issues still remaining to be solved, including the analysis of reflexive and multiple relations. In conventional social network analysis, the main focus has been on the structural aspect of networks[1], while the reflexive feature of self-loops (dialog with the self) and multiple dyadic relations (extended dialog with alters) are largely ignored.

Self-loops and multiple links can be observed, however, by monitoring networks of naturally emerging human behavior such as the networks in Wikipedia editing. Taking the English Wikipedia as an example, we found that the number of self-loops is more than half of the total edits and the number of multiple dyadic edits is more than 14% of the total edits. Hence, network analysis excluding this big portion of data will grossly distort the editing behavior of Wikipedians, easily amounting to up to 80% of the total interaction.

**BEHAVIORAL NETWORK ANALYSIS**

In this study, we focus on the behavioral aspects of social network analysis that may provide insights into the activities of actors in the network. The behavioral networks are prevalent in precedence relationships such as teamwork, group discussion, decision making, blogs, Wikipedia, and even team sports such as soccer for example. In teamwork, the repeated work of a person consists of self-loops, and repeated dyadic collaboration between two persons illustrates multiple links. Likewise, in group discussion, repeated talk of a single person is an example of self-loops and repeated interaction between two persons is a multiple link. In the case of soccer, repeated dribbling by a player corresponds to self-loops and repeatedly passing the ball between two players corresponds to multiple links. In collaborative editing of Wikipedia, we observe many self-loop and multiple link edits. Because of the data availability and reliability, the Wikipedia editing network is chosen to illustrate the behavioral analysis in this study.

For the behavioral analysis of the network, we defined measures to extract the behavioral pattern of actors including self-loop ratio, multiple-link ratio, speed, active ratio, anonymity ratio, Gini coefficient and Pareto ratio, as given in Table 1. Since most of the networks emerging in a natural context follow the power law with the scale-free property, ordinary scale measures are meaningless. As an illustration, the speed of editing is not measured by the ordinary measure of average editing intervals but by the ratio of the number of edits in a certain time frame.

**Behavioral Measures**

A self-loop is a circling reflexive link, which, in Wikipedia's case, is repeated circular editing by a single person before anyone else edits; the self-loop ratio is defined as the ratio of the number of self-loops to the total number of links. The reasons for self-loop editing may be multi-faceted; some people may do it without deliberate thinking before posting an idea, some may prefer to just post fast small increments, while some may find the article too interesting to pause for refinement. Overall, the act of self-loop editing may reflect the behavioral pattern of outgoingness, impatience and imprecision. The multiple-link ratio is the ratio of the number of repeated link pairs between the nodes to the total number of links, which implies dyadic collaboration.

As explained previously, speed is defined as the ratio of the number of edits in a certain time interval, e.g. one hour, to the total number of edits, which may express the behavior of quickness. The active ratio is defined as the ratio of the number of editors with more than a certain number of edits, e.g. 10, to the total number of editors, which may represent the behavior of active participation or voluntary contribution. The anonymity ratio is the ratio of the number of anonymous editors without registration to the total number of editors, which may express the behavior of self-concealment and closedness. The Gini coefficient is a measure of the inequality of a distribution of the number of edits among the editors.

## CULTURAL INTERPRETATIONS

Using the newly defined behavioral measures, we analyzed the behaviors of Wikipedia editors. For the analysis, we collected editing data of Wikipedia in 3 languages; Japanese, Chinese and Korean.

In order to further explore online culture, cultural dimensions are explored and postulated from behavioral patterns which are extracted by combining the structural and behavioral measures. Also, grouping is performed on the cultural dimensions based on Ward's minimum-variance cluster analysis, a hierarchical cluster analysis to show the proximity, and the Silhouette width criterion where the maximum width indicates the optimal number of clusters.

Considering the diverse cultural dimensions from the previous culture studies, the online cultural dimensions are explored and postulated by combining structural and behavioral measures as follows; collectivism vs. individualism, extraversion vs. introversion, boldness vs. deliberation, and egalitarianism vs. inegalitarianism. Collectivism is explored by a structural measure of the clustering coefficient complemented by the collaborative behavior observed by the multiple link ratio, i.e. dyadic collaboration. Extraversion is assumed to be manifested by the combination of outgoing and revealing behavior, i.e. as represented by the self-loop ratio and (opposite) anonymity ratio. Boldness may be partially related with extraversion and is postulated from outgoing and fastness behavior, i.e. as represented by the combined the self-loop ratio and speed. Egalitarianism is assumed to be manifested by the combination of equal and non-concentrated contribution behavior, i.e. as represented by the Gini coefficient and the Pareto ratio.

## DISCUSSION

The most collective country is Korea then China, and Japan is the least collective or the most individualistic. Korea is the most extraverted then China, and Japan. Again, Korea is the boldest then China, and Japan. Japan showed the least bold or highly deliberate. Japan exhibited the highest egalitarianism, and Korea and China showed the low egalitarianism.

Behavioral network analysis can be a powerful tool for the analysis of human behavior related to performance. For teamwork as an example, structural aspects of degree, density and centrality may not be sufficient to explain the success or failure of a team but the behavior of team members involving traits such as patience, collaboration, speed, active participation, openness and equality may be a better indicator to explain the performance of a team.

**Acknowledgements:** Supported by a Korea Research Foundation Grant, funded by the Korean Government (MEST)(KRF-2008-220-B00007).